\documentstyle[11pt,epsfig]{article}

\makeatletter

\def\@aabuffer{}
\def\author #1{\expandafter\def\expandafter\@aabuffer\expandafter
{\@aabuffer \small\rm      #1\relax \par}}
\def\address#1{\expandafter\def\expandafter\@aabuffer\expandafter
{\@aabuffer \small\it #1\relax 
\par\vspace{1em}}}

\def\maketitle{
\begin{center}
   {\bf \@title \par}       
   \vskip 2em                      % Vertical space after title.
   \@aabuffer\relax
\end{center} \par
\gdef\@aabuffer{}
}

\def\abstracts#1{
\begin{center}
{\begin{minipage}{5.2truein}
                 \footnotesize
                 \parindent=0pt #1\par
                 \end{minipage}}\end{center}
                 \vskip 2em \par}

\fussy
\def\section{\@startsection {section}{1}{\z@}{-3.5ex plus -1ex minus 
    -.2ex}{2.3ex plus .2ex}{\bf }}
\def\subsection{\@startsection{subsection}{2}{\z@}{-3.25ex plus -1ex minus 
   -.2ex}{1.5ex plus .2ex}{\it }}

\def\@makefnmark{{$\!^{\@thefnmark}$}}

\renewenvironment{thebibliography}[1]
	{\begin{list}{\arabic{enumi}.}
	{\usecounter{enumi}\setlength{\parsep}{0pt}
	 \setlength{\itemsep}{0pt} 
         \settowidth
	{\labelwidth}{#1.}\sloppy}}{\end{list}}

\newcounter{arabiclistc}

\def\@citex[#1]#2{\if@filesw\immediate\write\@auxout
	{\string\citation{#2}}\fi
\def\@citea{}\@cite{\@for\@citeb:=#2\do
	{\@citea\def\@citea{,}\@ifundefined
	{b@\@citeb}{{\bf ?}\@warning
	{Citation `\@citeb' on page \thepage \space undefined}}
	{\csname b@\@citeb\endcsname}}}{#1}}

\newif\if@cghi
\def\cite{\@cghitrue\@ifnextchar [{\@tempswatrue
	\@citex}{\@tempswafalse\@citex[]}}
\def\citelow{\@cghifalse\@ifnextchar [{\@tempswatrue
	\@citex}{\@tempswafalse\@citex[]}}
\def\@cite#1#2{{$\!^{#1}$\if@tempswa\typeout
	{IJCGA warning: optional citation argument 
	ignored: `#2'} \fi}}

\def\baselinestretch{1.0}
\ifx\selectfont\undefined
\@normalsize\else\let\glb@currsize=\relax\selectfont
\fi

\ifx\selectfont\undefined
\def\@singlespacing{%
\def\baselinestretch{1}\ifx\@currsize\normalsize\@normalsize\else\@currsize\fi%
}
\else
\def\@singlespacing{\def\baselinestretch{1}\let\glb@currsize=\relax\selectfont}
\fi

\long\def\@makecaption#1#2{
   \vskip 10pt 
   \setbox\@tempboxa\hbox{\footnotesize #1: #2}
   \ifdim \wd\@tempboxa >\hsize   % IF longer than one line:
       \leftskip 0pt plus 1fil 
       \rightskip 0pt plus -1fil 
       \parfillskip 0pt plus 2fil 
       \footnotesize #1: #2\par   %   THEN set as ordinary paragraph.
     \else                        %   ELSE  center.
       \hbox to\hsize{\hfil\box\@tempboxa\hfil}  
   \fi}

\makeatother

\def\be{\begin{equation}}
\def\ee{\end{equation}}
\def\bea{\begin{eqnarray}}
\def\eea{\end{eqnarray}}

\begin{document}
\vspace*{4cm}
\title{VARIATIONS OF $G$ AND SEE PROJECT}

\author{ V.N. MELNIKOV }

\address{Centre for Gravitaion and Fundamental Metrology, VNIIMS \\
3-1 M.Ulyanovoy Street, Moscow 117313, Russia and \\ Institute of
Gravitation and Cosmology, Peoples' Frienship University of Russia \\
E-mail: melnikov@rgs.phys.msu.su}

\maketitle\abstracts{
Problems of absolute $G$ measurements, its temporal and range variations
>from both experimental and theoretical points of view are discussed, and
a new universal space project for measuring $G$, $G(r)$ and $G$-dot
promising an improvement of our knowledge of these quantities by 2-3
orders is advocated.}

%%%%%%%%%%%%%%%%%%%%%%%%%%%%%%%%%%%%%%%%%%%%%%%%%%%%%%%%%%%%%%%%%%%%%%%%%%%
\section{Introduction}
%%%%%%%%%%%%%%%%%%%%%%%%%%%%%%%%%%%%%%%%%%%%%%%%%%%%%%%%%%%%%%%%%%%%%%%%%%%

Among fundamental physical constants Newton's gravitational constant $G$ (as
well as other fundamental macroscopic parameters H - the Hubble constant,
$\rho$, or $\Omega$, - mean density of the Universe, $\Lambda$ - the
cosmological constant) is known with the least accuracy.\cite{1,2} Moreover,
there are other problems related to its possible variations with
time\cite{13,14} and range\cite{4} coming mainly from predictions of unified
models of the four known physical interactions.

Here we dwell upon the problems of absolute $G$ measurements, its
temporal and range variations from both experimental and theoretical points
of view and advocate a new universal space project for measuring $G$, $G(r)$
and $G$-dot promising an improvement of our knowledge of these quantities
by 2-3 orders.

Why are we interested in an absolute value of $G$? It is known in
celestial mechanics that we can determine only the product $GM$, where $M$
is the mass of a body. ($GM$ is known now with the accuracy $10^{-9}$ which
is much better than $10^{-3}$ for $G$ and correspondingly for masses of
planets.) The knowledge of this product is enough for solving most problems
in celestial mechanics and space dynamics.  But there are other areas where
we need separate values of $G$ and $M$.  First, we need to know much better
masses of planets for construction of their precise models.  Second, $G$
enters indirectly in some basic standards and is necessary for conversion
>from mechanical to electromagnetic units, for calibration of gradiometers
etc.  More precise value of $G$ will be important for future dicrimination
between unified (GUT, supergravity, strings, M-theory etc.) models as they
usually predict certain relations between $G$ and other fundamental
constants.\cite{1,5,10,11,15}

What is the experimental situation with $G$?

\newpage

\section{Problem of Stability of $G$}

\subsection{Absolute $G$ measurements}

The value of the Newton gravitational constant $G$ as adopted by CODATA
in 1986 is based on the Luther and Towler measurements of 1982.

Even at that time other existing on 100ppm level measurements deviated
>from this value more than their uncertainties.\cite{6} During recent years
the situation, after very precise measurements of $G$ in Germany and New
Zealand, became much more vague. Their results deviate from the official
CODATA value drastically.

As it is seen from the most recent data announced in November 1998 at
the Cavendish conference in London the situation with terrestrail
absolute $G$ measurements is not improving.
The reported values for $G$ (in units of $10^{11}$ SI) and their estimated
error in ppm are as follows:

\begin{center}
\medskip
\begin{tabular}{lll}
Fitzgerald and Armstrong & 6.6742 & 90 ppm \\
(NZ)                     & 6.6746 & 134 \\
Nolting et al. (Zurich)	& 6.6749 & 210 \\
Meyer et al. (Wupperthal) & 6.6735 & 240 \\
Karagioz et al. (Moscow) & 6.6729 & 75 \\
Richman et al. & 6.683 & 1700 \\
Schwarz et al. & 6.6873 & 1400 \\
CODATA (1986, Luther) & 6.67259 & 128
\end{tabular}
\medskip
\end{center}

This means that either the limit of terrestrial accuracies is reached or
we have some new physics entering the measurement procedure.\cite{7,8,15}
The first means that we should address to space experiments to measure $G$
\cite{9} and the second means that a more thorough study of theories
generalizing Einstein's general relativity is necessary.

\subsection{Data on temporal variations of G}

Dirac's prediction based on his Large Numbers Hypothesis is $\dot{G}/G =
(-5)10^{-11}$ $year^{-1}$. Other hypotheses and theories, in particular
some scalar-tensor or multidimensional ones, predict these
variations on the level of $10^{-12}-10^{-13}$ per year. As to experimental
or observational data, the results are rather inconclusive. The most
reliable ones are based on Mars orbiters and landers (Hellings, 1983)
and on lunar laser ranging (Muller et al., 1993; Williams et al., 1996).
They are not better than $10^{-12}$ per year.\cite{7,15} Here are some
data on $\dot G$:

\begin{center}
\medskip
\begin{tabular}{ll}
1. Van Flandern, 1976-1981: & $\dot G/G=-5\cdot10^{-11}y^{-1}$ \\
\ \ \ (ancient eclipses) & \\
2. Hellings, 1983-1987: & $|\dot G/G|<5\cdot10^{-12}y^{-1}$ \\
\ \ \ (Viking) & \\
3. Reasenberg, 1987: & $|\dot G/G|<5\cdot10^{-11}y^{-1}$ \\
\ \ \ (Viking) & \\
4. Acceta et al., 1992: & $|\dot G/G|<10^{-12}y^{-1}$ \\
\ \ \ (Nucleosyntheses) & \\
5. Anderson et al., 1992: & $|\dot G/G|\le2\cdot10^{-12}y^{-1}$ \\
\ \ \ (ranging to Mercury and Venus) & \\
6. Muller et al., 1993: & $|\dot G/G|\le5\cdot10^{-13}y^{-1}$ \\
\ \ \ (lunar laser ranging) & \\
7. Kaspi et al., 1994: & $|\dot G/G|\le5\cdot10^{-12}y^{-1}$ \\
\ \ \ (timing of pulsars) & \\
8. Williams et al., 1996: & $|\dot G/G|\le8\cdot10^{-12}y^{-1}$ \\
\ \ \ (lunar laser ranging) &
\end{tabular}
\medskip
\end{center}

Here once more we see that there is a need for corresponding theoretical and
experimental studies. Probably, future space missions to other planets will
be a decisive step in solving the problem of temporal variations of $G$ and
determining the fate of different theories which predict them as the larger
is the time interval between successive measurements and, of course, the
more precise they are, the more stringent results will be obtained.

\subsection{Nonnewtonian interactions (EP and ISL tests)}

Nearly all modified theories of gravity and unified theories predict also
some deviations from the Newton law (ISL) or composite-dependent violation
of the Equivalence Principle (EP) due to appearance of new possible
massive particles (partners).\cite{4}

In the Einstein theory $G={\rm const}$. If $G=G(t)$ is possible, then, from
the relativistic point of view $G\to G(t,r)$. In more detail: Einstein's
theory corresponds to massless gravitons which are mediators of
the gravitational interaction, obey 2nd order differential equations and
interact with matter with a constant strength $G$.

Any violation of these conditions leads to deviations from the Newton Law.
Here are some classes of theories (generalized gravitational and inified
models) which exibit such deviations.\cite{}

1. Massive gravitons: theories with $\Lambda$ and bimetric.

2. Effective $G(x,t)$: scalar-tensor theories.

3. Theories with torsion.

4. Theories with higher derivatives (4th order etc.). Here massive modes
appear:  short and long range forces.

5. Other mediators besides gravitons (partners) appear: $Super Gravity$,
$Super Strings$, $M-theory$ etc. (massive).

6. Theories with nonlinearities induced by any of the known interactions
(electromagnetic or gravitational or other). Then, some mass of the mediators
appears.

7. Phenomenological models where the mechanism of deviation from the Newton
law is not known (fifth force or so). For describing the possible deviation
>from the Newton law the usual parametrization
$\Delta\sim\alpha{\rm e}^{-r/\lambda}$ of the Yukawa-type is used.

Experimental data exclude the existence of these particles at nearly all
ranges except less than a millimeter and  also at meters and hundreds of
meters ranges. Here are some estimations of masses, ranges and also
strengths for $G(r)$ predicted by various models.

1. Pseudoscalar particle leads to attraction between macro bodies with range
$2\cdot10^{-4}\,{\rm cm}<\lambda<20\,{\rm cm}$ (Moody, Wilizek, 1984), the
variable $\alpha$ (strength) from 1 to $10^{-10}$ in this range is predicted.

2. Supersymmetry: spin-1 partner of the massive spin-3/2 gravitino leads to
repulsion in the range: $\lambda\sim10^3$ km, $\alpha\sim10^{-13}$
(Fayet, 1986, 1990).

3. Scalar field to adjust $\Lambda$ (Weinberg, 1989): $m\le10^{-3}\,eV/c^2$
or range $\lambda\ge0.1$ mm. Another variant (Peccei, Sola, Wetterich, 1987)
leads to $\lambda\le10$ km (attraction).

4. Supergravity (Scherk, 1979); graviton is accompanied by a spin-1
graviphoton:  here a repulsion is predicted also.

5. Strings, p-branes, M-theory: dilaton (other scalar fields) and
antisymmetric tensor fields appear.

{\bf Conclusion:} there is no reliable theory or model of unified type, but
all predict new interactions of a non-Newtonian character (composition
dependent $EP$-violation or independent).

It is a real challange to experimental people!

It should be noted that the most recent result in the range of 20-500 m was
obtained by Achilli et al.\cite{12} They found a
deviation from the Newton law with the Yukawa potential strength $\alpha$
between 0,13 and 0,25.  Of course, these results need to be verified in
other independent experiments, probably in space ones.

\section{SEE Project}

We saw that there are three problems connected with $G$: absolute value
mesurements and possible variations with time and range. There is a
promissing new space experiment SEE - Satellite Energy Exchange\cite{9}
which adresses to all these problems and may be more effective in solving
them than other laboratory or space experiments.

We studied some aspects of the SEE-Project:\cite{17}

1. Wide range of trajectories with the aim of finding optimal
ones:

-- circular in spherical field;

-- circular in spherical field $+$ earth quadrupole modes;

-- elliptic, with $e\le0.05$.

2. Estimations of other celistical bodies influence.

3. Estimations of relative influence of trajectories to $\delta G$,
$\delta\alpha$.

4. Modelling measurement procedure of $G$ and $\alpha$.

5. Estimations of some sources of errors:

-- radial oscilations of the shepherd's surface;

-- longitudal oscilations of the capsule;

-- transversal oscilations of the capsule;

-- shepherd nonsphesicity;

-- limits on shepherd $J_2$.

6. Error budgets for $G$, $G$-dot and $G(r)$. The general conclusion is that
the Project SEE may improve our knoweledge of $G$, limits on $G$-dot and
$G(r)$ by 2-3 orders of magnitude.

7. Variation of the SEE method -- trajectories near libration points.

8. Different altitudes up to ISS (500 km), short capsule up to 5 m
(instead of original 15-20 m).

{\bf General conclusion:} it is possible to improve $(G,\alpha)$ by
2-3 orders at a range of 1-100 m.

%%%%%%%%%%%%%%%%%%%%%%%%%%%%%%%%%%%%%%%%%%%%%%%%%%%%%%%%%%%%%%%%%%%%%%%%%%%
\section*{Acknowledgments}
%%%%%%%%%%%%%%%%%%%%%%%%%%%%%%%%%%%%%%%%%%%%%%%%%%%%%%%%%%%%%%%%%%%%%%%%%%%

The author is grateful to NASA and the University of Tennessee for the
partial support of this work and also to the organizers of the Rencontre du
Moriond-99 for their hospitality.

%%%%%%%%%%%%%%%%%%%%%%%%%%%%%%%%%%%%%%%%%%%%%%%%%%%%%%%%%%%%%%%%%%%%%%%%%%%
\section*{References}
%%%%%%%%%%%%%%%%%%%%%%%%%%%%%%%%%%%%%%%%%%%%%%%%%%%%%%%%%%%%%%%%%%%%%%%%%%%

\end{document}